\documentclass{ws-mpla}

\begin{document}

\markboth{Yuri Shtanov, Alexander Viznyuk, and Luis Norberto Granda}
{Asymmetric Embedding in Brane Cosmology}

\catchline{}{}{}{}{}
%
%%%%%%%%%%%%%%%%%%%%%%%%%%%%%%%%%%%%%%%%%%%%%%%%%%%%%%%%%%%%%%%%%%%%

\title{ASYMMETRIC EMBEDDING IN BRANE COSMOLOGY}

\author{YURI SHTANOV$^*$ and ALEXANDER VIZNYUK$^\dagger$}
%\address{Bogolyubov Institute for Theoretical Physics, Kiev 03680, Ukraine \\
%shtanov@bitp.kiev.ua}

\address{Bogolyubov Institute for Theoretical Physics, Kiev 03680, Ukraine \\
$^*$shtanov@bitp.kiev.ua \\ $^\dagger$viznyuk@bitp.kiev.ua}

\author{LUIS NORBERTO GRANDA}

\address{Departamento de F\'{\i}sica, Universidad del Valle, A.A.~25360, Cali, Colombia \\
ngranda@univalle.edu.co}

\maketitle

\pub{}{}

\begin{abstract}
We derive a system of cosmological equations for a braneworld with induced
curvature which is a junction between several bulk spaces. The permutation
symmetry of the bulk spaces is not imposed, and the values of the fundamental
constants, and even the signatures of the extra dimension, may be different on
different sides of the brane.  We then consider the usual partial case of two
asymmetric bulk spaces and derive an exact closed system of scalar equations on
the brane. We apply this result to the cosmological evolution on such a brane
and describe its various partial cases.

\end{abstract}

\ccode{PACS Nos.: 04.50.+h, 98.80.Es}

\section{Introduction}

The idea that our four-dimensional world can be described as a timelike hypersurface
(brane) embedded in or bounding a five-dimensional manifold continues to be in the focus
of modern investigations.  Especially this concerns cosmological braneworld solutions
which exhibit many interesting and unusual properties (see
Refs.~\refcite{Maartens}--\refcite{Koyama} for recent reviews). Theories with the
simplest generic action involving scalar-curvature terms both in the bulk and on the
brane allow for the possibilities of superacceleration of the universe
expansion,\cite{SS1,SS2} cosmological loitering even in a spatially flat
universe,\cite{loiter} and ``cosmic mimicry,''\cite{mimicry} which is characterized by
the property that the braneworld model at low redshifts is virtually indistinguishable
from the LCDM ($\Lambda$ + Cold Dark Matter) cosmology but has renormalized value of the
cosmological density parameters. The possibility of explaining other dark-matter
phenomena in such models is discussed in Ref.~\refcite{VS}.

A vast majority of papers on braneworld cosmology are dealing with the
situation of the $Z_2$ symmetry of reflection with respect to the brane, which
is equivalent to the brane being the boundary of the bulk.  The present letter
is devoted to the investigation of solutions where such symmetry is not
imposed.

Braneworld cosmologies without $Z_2$ symmetry of reflection were a subject of
investigation, for example, in Refs.~\refcite{Kraus}--\refcite{DVDP}. In these papers,
however, the curvature term in the action on the brane was absent. The braneworld model
without $Z_2$ symmetry was studied in Ref.~\refcite{KLM} in the presence of the curvature
term on the brane and with equal bulk cosmological and gravitational constants on either
side of the brane. In Ref.~\refcite{Shtanov1}, the $Z_2$-asymmetric model was considered
in the braneworld theory with curvature term on the brane and with different cosmological
constants in the bulk but with equal bulk gravitational constants on either sides of the
brane. In a recent paper,\cite{KK} an asymmetric braneworld cosmology was under
consideration in the model without the induced-curvature term on the brane but with
different bulk cosmological and gravitational constants on either side of the brane.

In this letter, we would like to consider the most generic braneworld model in which the
values of the bulk fundamental (gravitational and cosmological) constants are different
on the two sides of the brane and, in addition to this, the signatures of the extra
dimension on the two sides of the brane may be arbitrary (and different).  Note that the
$Z_2$-symmetric braneworld cosmology with timelike extra dimension was previously
considered in Refs.~\refcite{Kofinas}, \refcite{CK}--\refcite{SS3}.

Before considering the asymmetric case with a brane separating between two bulk
spaces, we consider a generalization in which a brane is a multivolume
junction, i.e., a boundary of $N$ bulk spaces with natural junction conditions.
The theory of such spaces was initiated in Ref.~\refcite{FS}. In the next
section, we derive a system of cosmological equations for this configuration
without assuming any symmetry between the bulk spaces. In particular, the
values of the fundamental constants, and even the signatures of the extra
dimension, may be different on different sides of the brane. After that, we
study in more detail the usual case of two bulk spaces.

\section{A brane as a multivolume junction}

\begin{figure}
\center{\includegraphics[width=.55\textwidth]{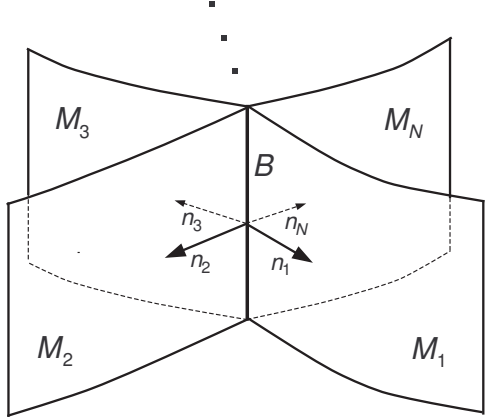}} \caption{Brane
${\mathcal B}$ as a junction between the bulk spaces ${\mathcal M}_1,\ldots,
{\mathcal M}_N$.} \label{fig:brane}
\end{figure}

Our starting setup is a four-dimensional hypersurface (brane) ${\mathcal B}$
which is simultaneously a boundary of $N$ five-dimensional Riemannian manifolds
(bulk spaces) ${\mathcal M}_1, \ldots, {\cal M}_N$ and has nondegenerate
Lorentzian induced metric (see Fig.~\ref{fig:brane}).
The primary junction condition which makes the brane the boundary of all bulk
spaces is that the induced metric is one and the same in all ${\mathcal M}_I$,
$I = 1, \ldots, N$. We assume that the bulk gravitational and cosmological
constants on different sides of the brane can be different. The action of the
theory has the form
\begin{equation} \label{action}
S = \sum_{I = 1}^N M_I^3  \left[ \int_{{\mathcal M}_I} \left({\mathcal R}_I - 2
\Lambda_I \right) - 2 \epsilon_I \int_{{\mathcal B}} K_I \right] +
\int_{{\mathcal B}} \left( m^2 R - 2 \sigma \right) + \int_{{\mathcal B}} L
(h_{ab}, \phi) \, .
\end{equation}
Here, ${\mathcal R}_I$ is the scalar curvature of the five-dimensional metric
$g^I_{ab}$ on ${\mathcal M}_I$, and $R$ is the scalar curvature of the induced
metric $h_{ab} = g_{ab} - n^I_a n^I_b$ on ${\mathcal B}$, where $n_I^a$ is the
vector of the inner unit normal to the brane.\footnote{Although the vector of
inner unit normal $n_I^a$ is different in each of the bulk parts ${\mathcal
M}_I$, the induced metric $h_{ab}$ is one and the same.} The quantity $K_I =
K^I_{ab} h^{ab}$ is the trace of the symmetric tensor of extrinsic curvature
$K^I_{ab}$ of ${\mathcal B}$ in ${\mathcal M}_I$. The parameter $\epsilon_I =
1$ if the signature of the corresponding bulk part ${\mathcal M}_I$ is
Lorentzian, so that the extra dimension is spacelike, and $\epsilon_I = -1$ if
its signature is $(-,-,+,+,+)$, so that the extra dimension is timelike. The
symbol $L (h_{ab}, \phi)$ denotes the Lagrangian density of the
four-dimensional matter fields $\phi$ the dynamics of which is restricted to
the brane ${\mathcal B}$ so that they interact only with the induced metric
$h_{ab}$. All integrations over ${\mathcal M}_I$ and over ${\mathcal B}$ are
taken with the corresponding natural volume elements. The symbols $M_I$, $I =
1, \ldots, N$, denote the Planck masses of the corresponding spaces,
$\Lambda_I$, $I = 1, \ldots, N$, are the corresponding five-dimensional
cosmological constants, and $m$ and $\sigma$ are the Planck mass and tension of
the brane, respectively.

In this letter, we systematically use the notation and conventions of
Ref.~\refcite{Wald}. In particular, we use the one-to-one correspondence
between tensors in ${\mathcal B}$ and tensors in ${\mathcal M}$ which are
invariant under projection to the tangent space to ${\mathcal B}$, i.e.,
tensors $T^{a_1 \cdots a_k}{}_{b_1 \cdots b_l}$ such that
\begin{equation}
T^{a_1 \cdots a_k}{}_{b_1 \cdots b_l} = h^{a_1}{}_{c_1} \cdots
h^{a_k}{}_{c_k} h_{b_1}{}^{d_1} \cdots h_{b_l}{}^{d_l} T^{c_1 \cdots
c_k}{}_{d_1 \cdots d_l} \, .
\end{equation}

Variation of action (\ref{action}) gives the equation of motion in the
five-dimensional bulk parts ${\mathcal M}_I$:
\begin{equation} \label{bulk}
{\mathcal G}^I_{ab} + \Lambda_I g^I_{ab} = 0 \, , \quad I = 1,\ldots,N \, ,
\end{equation}
and on the brane ${\mathcal B}$:
\begin{equation} \label{brane}
m^2 G_{ab} + \sigma h_{ab} = \sum_{I = 1}^N \epsilon_I M_I^3\, S^I_{ab} +
\tau_{ab} \, ,
\end{equation}
where ${\mathcal G}^I_{ab}$ and $G_{ab}$ are the Einstein's tensors of the
corresponding spaces, $S^I_{ab} \equiv K^I_{ab} - K^I h_{ab}$ is formed from
the tensor of extrinsic curvature, and $\tau_{ab}$ denotes the four-dimensional
stress--energy tensor of matter on the brane.  It is the presence of the
tensors $S^I_{ab}$ in the equation of motion (\ref{brane}) that makes the inner
dynamics on the brane ${\mathcal B}$ unusual.\footnote{Those interested in the
derivation of (\ref{brane}) may look into the appendices of
Refs.~\refcite{FS,Shtanov}.}

The Codazzi relations on the brane, in view of the bulk equation (\ref{bulk}),
read
\begin{equation} \label{conserve-s}
D^a S^I_{ab} = {\mathcal R}^I_{cd} n_I^d h^c{}_b = 0 \, , \quad I = 1, \ldots,
N \, ,
\end{equation}
where $D^a$ is the (unique) derivative on the brane ${\mathcal B}$ associated
with the induced metric $h_{ab}$.  Equation (\ref{brane}) then implies the
relation
\begin{equation} \label{conserve}
D_a \tau^a{}_b = 0 \, .
\end{equation}
Thus, the four-dimensional stress--energy tensor is covariantly conserved on
the brane, which is a consequence of the absence of matter in the bulk.

Generalizing the procedure of Refs.~\refcite{BDL,SMS} to the case of $N$ bulk
spaces, we consider the Gauss identity:
\begin{equation} \label{gauss}
R_{abcd} = h_a{}^f h_b{}^g h_c{}^k h_d{}^j\, {\mathcal R}^I_{fgkj} + \epsilon_I
\left( K^I_{ac} K^I_{bd} - K^I_{bc} K^I_{ad} \right) \, , \quad I = 1,\ldots,N
\, .
\end{equation}
Contracting this relation and taking into account Eq.~(\ref{bulk}), one obtains
the equation
\begin{equation} \label{constraint}
\epsilon_I \left( R - 2 \Lambda_I \right) + K^I_{ab} K_I^{ab} - K_I^2 \equiv
\epsilon_I \left( R - 2 \Lambda_I \right) + S^I_{ab} S_I^{ab} - \frac13 S_I^2 =
0 \, , \quad I = 1,\ldots,N \, ,
\end{equation}
which is valid on all sides of the brane, and which we expressed in terms of
$S^I_{ab} = K^I_{ab} - h_{ab} K^I$ and $S^I = h^{ab} S^I_{ab}$. This is the
well-known constraint equation on the brane from the viewpoint of the
gravitational dynamics in the five-dimensional bulk.

Our aim is to derive the resulting cosmological equation on the brane. One
could solve this problem by considering embedding of the brane in the bulk
spaces under consideration and calculating the corresponding extrinsic
curvatures.  However, in a cosmological setup, one can integrate the constraint
equations (\ref{constraint}) directly on the brane.  First, we note that, for
any tensor on the brane $T_{ab}$ which is covariantly conserved, i.e., $D_a
T^a{}_b = 0$, in the cosmological setup
\begin{equation} \label{T}
T^0{}_0 = \beta (t) \, ,  \qquad T^\mu{}_\nu = \delta^\mu{}_\nu q(t) \, , \quad
\mu, \nu = 1, 2, 3 \, ,
\end{equation}
one can easily verify the following relation:
\begin{equation} \label{relation}
T_{ab} T^{ab} - \frac13 T^2 = \frac{1}{3a^4 H} \frac{d}{dt} \left(a^4 \beta^2
\right) \, .
\end{equation}
In view of the conservation equation, the function $q(t)$ in (\ref{T}) is
uniquely expressed through $\beta (t)$. Then, setting
\begin{equation}
S_I^0{}_0 = - 3 \beta_I (t) \, , \quad I = 1, \ldots, N \, ,
\end{equation}
and using property (\ref{relation}), valid for all these tensors in view of the
conservation equation (\ref{conserve-s}), we can integrate
Eq.~(\ref{constraint}) with the result
\begin{equation} \label{betai}
\epsilon_I \left( H^2 + \frac{\kappa}{a^2} - \frac{\Lambda_I}{6} \right) =
\beta_I^2 + \frac{\epsilon_I C_I}{a^4} \, , \quad I = 1, \ldots, N \, ,
\end{equation}
where $C_I$ are integration constants.  The zero--zero component of
Eq.~(\ref{brane}) gives
\begin{equation}
H^2 + \frac{\kappa}{a^2} = \frac{\rho + \sigma}{3 m^2} + \frac{1}{m^2} \sum_{I
= 1}^N \epsilon_I  M_I^3 \beta_I \, .
\end{equation}
Substituting $\beta_I$ found from Eq.~(\ref{betai}) into this equation, we
obtain our main result:
\begin{equation} \label{junction}
H^2 + \frac{\kappa}{a^2} = \frac{\rho + \sigma}{3 m^2} + \frac{1}{m^2} \sum_{I
= 1}^N \zeta_I  M_I^3 \left[\epsilon_I \left(H^2 + \frac{\kappa}{a^2} -
\frac{\Lambda_I}{6} - \frac{C_I}{a^4} \right) \right]^{1/2} \, ,
\end{equation}
where $\zeta_I = \pm 1$ corresponds to the possibility of different signs in
the solution for $\beta_I$ from Eq.~(\ref{betai}).  Physically, these signs
correspond to the two possible ways of bounding each of the spaces ${\mathcal
M}_I$ by the brane.

The integration constants $C_I$ generalize the so-called ``dark radiation''
contribution to the dynamics of the brane.  In the case where the brane is a
boundary of $N$ independent bulk spaces, there are exactly $N$ such independent
integration constants.  If nonzero, they reflect the existence of black holes
in the corresponding bulk spaces.

\section{The case of two bulk spaces}

In this section, we consider in detail the case where the brane is just
embedded in one bulk space, so that $N = 2$, i.e., there are only two ``sides''
of the brane. Equation (\ref{brane}) involves the tensors $S^I_{ab}$ which are
constructed from the tensors of extrinsic curvature of the brane, so this
equation is not closed with respect to the intrinsic evolution on the brane.
However, using (\ref{constraint}), it is possible to obtain a system of scalar
equations which involves only four-dimensional fields on the brane. Following,
with slight modifications, the procedure first adopted in
Ref.~\refcite{Shtanov1} for the particular case $M_1 = M_2$ and $\epsilon_1 =
\epsilon_2 = 1$, we introduce the tensors
\begin{equation}
\Sigma_{ab} = \epsilon_{1}M_{1}^{3} S^{(1)}_{ab} + \epsilon_{2}M_{2}^{3}
S^{(2)}_{ab} \, , \qquad \Delta_{ab} = \epsilon_{1}M_{1}^{3}S^{(1)}_{ab} -
\epsilon_{2}M_{2}^{3} S^{(2)}_{ab} \, .
\end{equation}
Rewriting (\ref{constraint}) in terms of $\Sigma_{ab}$ and $\Delta_{ab}$, we
easily obtain the following closed system of gravitational equations on the
brane:
\begin{eqnarray}
\left(\Sigma_{ab} \Sigma^{ab} - \frac13 \Sigma^2 \right) +
\left(\Delta_{ab} \Delta^{ab} - \frac13 \Delta^2 \right) \nonumber \\
{} + 2 \left(\epsilon_1 M_1^6 + \epsilon_2 M_2^6 \right) R - 4 \left(\epsilon_1
M_1^6 \Lambda_1 +\epsilon_2 M_2^6 \Lambda_2\right) = 0 \, ,
\label{constraint-new}
\end{eqnarray}
\begin{equation} \label{orthog}
\left (\Sigma_{ab} \Delta^{ab} - \frac13 \Sigma \Delta\right) +
\left(\epsilon_1 M_1^6-\epsilon_2 M_2^6\right)R - 2\left(\epsilon_1 M_1^6
\Lambda_1 -\epsilon_2 M_2^6 \Lambda_2\right)=0 \, ,
\end{equation}
\begin{equation} \label{conserve-new}
D_a \Delta^a{}_b = 0 \, ,
\end{equation}
where $\Sigma = \Sigma_{ab} h^{ab}$, $\Delta = \Delta_{ab} h^{ab}$, and
$\Sigma_{ab}$ is given by
\begin{equation} \label{s-new}
\Sigma_{ab} =   m^2 G_{ab} + \sigma h_{ab} - \tau_{ab}
\end{equation}
in view of (\ref{brane}).

This system of equations is to be solved for the metric and matter fields and
for the symmetric tensor field $\Delta_{ab}$ on the brane.  It constitutes the
main system of closed scalar equations on the brane, which arises in the
absence of any information and/or boundary conditions for the brane--bulk
system.  For the particular case $M_1 = M_2$ and $\epsilon_1 = \epsilon_2 = 1$,
this system of equations was first obtained in Ref.~\refcite{Shtanov1}.

In what follows, we consider the cosmological implications of system
(\ref{constraint-new})--(\ref{s-new}) for the homogeneous and isotropic
cosmological model with the cosmological time $t$, scale factor $a (t)$, energy
density $\rho (t)$, and pressure $p (t)$. Under these conditions, we set the
tensor $\Delta_{ab}$ to be homogeneous and isotropic as well:
\begin{equation}
\Delta^0{}_0 = -\beta (t) \, , \qquad \Delta^\mu{}_\nu = \delta^\mu{}_\nu q (t)
\, , \quad \mu, \nu = 1,2,3 \, .
\end{equation}

Using (\ref{conserve}) and (\ref{conserve-new}), it is easy to calculate the
quantities
\begin{equation}
\Sigma_{ab} \Sigma^{ab} - \frac13 \Sigma^2 = - \frac{1}{3a^4 H} \frac{d}{dt}
\left[ a^2 \left( 3 m^2 \chi - \rho - \sigma \right) \right]^2 \, ,
\end{equation}
\begin{equation} \label{int-delta}
\Delta_{ab}\Delta^{ab} - \frac13 \Delta^2 = - \frac{1}{3a^4 H} \frac{d}{dt}
\left(a^4 \beta^2 \right) \, ,
\end{equation}
\begin{equation}
\Sigma_{ab}\Delta^{ab} - \frac13 \Sigma \Delta = -\frac{1}{3a^4 H} \frac{d}{dt}
\left[a^4 \beta \left( 3m^2 \chi - \rho - \sigma \right)\right]\,,
\end{equation}
where $H\equiv \dot{a}/a$ is the Hubble parameter,
\begin{equation}
\chi \equiv H^2 + \frac{\kappa}{a^2} \, ,
\end{equation}
and $\kappa = 0, \pm 1$ corresponds to the sign of the spatial curvature of the
model.

Since the expressions $a^3 \dot{a}$ and $a^3 \dot{a}R$ are also total
derivatives, equations (\ref{constraint-new}) and (\ref{orthog}) can be
integrated with the result
\begin{equation} \label{constraint-new-new}
m^4\left(\chi-\frac{\rho+\sigma}{3m^2}\right)^2+\frac{\beta^2}{9}=
2\left(\epsilon_1 M_1^6+\epsilon_2 M_2^6\right)\chi -
\frac{1}{3}\left(\epsilon_1 M_1^6 \Lambda_1+\epsilon_2 M_2^6
\Lambda_2\right)-\frac{C}{a^4}\, ,
\end{equation}
\begin{equation} \label{orthog-new}
\beta\left(3m^2 \chi-\rho-\sigma\right)=9\left(\epsilon_1 M_1^6-\epsilon_2
M_2^6\right)\chi-\frac32\left(\epsilon_1 M_1^6 \Lambda_1-\epsilon_2 M_2^6
\Lambda_2\right)-\frac{9E}{a^4}\, ,
\end{equation}
where $C$ and $E$ are integration constants.  Then, eliminating $\beta$ from
(\ref{orthog-new}) and using (\ref{constraint-new-new}), we finally obtain
\begin{eqnarray}
&{}&\displaystyle m^4 \left( H^2 + \frac{\kappa}{a^2} - \frac{\rho + \sigma}{3
m^2} \right)^2 - 2\left(\epsilon_1 M_1^6 + \epsilon_2 M_2^6\right)
\left(H^2+\frac{\kappa}{a^2}\right) \qquad\qquad\qquad\qquad\qquad
\nonumber \\
&{}&\displaystyle {} + \frac{1}{3} \left(\epsilon_1 M_1^6 \Lambda_1+\epsilon_2
M_2^6 \Lambda_2\right) + \frac{C}{a^4} \nonumber \\ &{}&\displaystyle {} +
\left[\frac{\left(\epsilon_1 M_1^6 - \epsilon_2 M_2^6 \right) \left( H^2 +
\kappa /a^2 \right) - \left( \epsilon_1 M_1^6 \Lambda_1-\epsilon_2 M_2^6
\Lambda_2 \right)/6 - E/a^4}{m^2 \left(H^2 + \kappa /a^2 \right) - (\rho +
\sigma)/ 3 }\right]^2=0 \, . \label{cosmol}
\end{eqnarray}

This is our main result as regards cosmology without $Z_2$ symmetry in the
bulk.  It could also be obtained directly from (\ref{junction}) with $I = 2$.
This equation in its general form is rather complicated, so it is useful to
consider some partial cases. The general case $\epsilon_1 = \epsilon_2 = +1$
will be investigated in more detail elsewhere.

\subsection{$\epsilon_1 = \epsilon_2 = +1$ and $\Lambda_1 = \Lambda_2$ or $M_1 = M_2$ }

We begin with the case $\Lambda_1 = \Lambda_2 = \Lambda$ and also set
$\epsilon_1 = \epsilon_2 = +1$. Then Eq.~(\ref{cosmol}) becomes\footnote{Here
and below, for notational simplicity, we omit the spatial curvature term
$\kappa / a^2$. It can easily be recovered by the substitution $H^2 \rightarrow
H^2+\kappa/a^2$ in all formulas.}
\begin{eqnarray} \label{equal lambda}
m^4 \left( H^2 - \frac{\rho + \sigma}{3 m^2} \right)^2 - 2 \left( M_1^6 +
M_2^6\right) \left(H^2-\frac{\Lambda}{6}\right) + \frac{C}{a^4}
\nonumber\\
{} +  \left[ \frac{ \left( M_1^6 - M_2^6 \right) \left( H^2 - \Lambda/6 \right)
- E / a^4}{m^2H^2 - (\rho + \sigma)/ 3} \right]^2 = 0 \, .
\end{eqnarray}

The last term on the right-hand side of this equation represents the difference
between our model and standard $Z_2$-symmetric cosmology. Setting additionally
$M_1 = M_2 = M$ and $E=0$, we obtain the well-known result of the
 $Z_2$-symmetric case:\cite{CH,Deffayet}
\begin{equation}\label{equal lambda and M}
m^4 \left( H^2 - \frac{\rho + \sigma}{3 m^2} \right)^2 = 4
M^6\left(H^2-\frac{\Lambda}{6}-\frac{C_1}{a^4}\right)\, , \qquad
C_1=\frac{C}{4M^6} \, .
\end{equation}

This equation can easily be solved with respect to the Hubble parameter, giving
two branches:\cite{SS1,SS2}
\begin{equation}\label{equal lambda and M resolved}
H^2 = \frac{\rho + \sigma}{3m^2} + \frac{2M^6}{m^4} \pm \frac{2M^6}{m^4} \left[
1 + \frac{m^4}{M^6} \left( \frac{\rho + \sigma}{3m^2} - \frac{\Lambda}{6} -
\frac{C_1}{a^4} \right)\right]^{1/2} \,.
\end{equation}

Setting $\epsilon_1 = \epsilon_2 = \epsilon$ and $M_1 = M_2 = M$ but different
$\Lambda_1$ and $\Lambda_2$ in Eq.~(\ref{cosmol}), we obtain
\begin{eqnarray}\label{equal M}
m^4 \left( H^2  - \frac{\rho + \sigma}{3 m^2} \right)^2 = 4 \epsilon M^6
\left(H^2 - \frac{\Lambda_1 + \Lambda_2}{12} - \frac{\epsilon C_1}{a^4} \right)
\nonumber \\ - \frac{M^{12}}{36} \left[ \frac{\Lambda_1 - \Lambda_2 +
E_1/a^4}{m^2 H^2 - (\rho + \sigma) / 3 } \right]^2 ,
\end{eqnarray}
where $C_1$ is defined in (\ref{equal lambda and M}), and
\begin{equation} \label{e1}
E_1 = \frac{6E}{M^6} \, .
\end{equation}
Equation (\ref{equal M}) coincides with the result of Ref.~\refcite{Shtanov1}.

\subsection{$\Lambda_1 = \Lambda_2$ and $M_1 = M_2$}

Another interesting possibility is to take all the constants of the theory
equal on the two sides of the brane (i.\,e., $\Lambda_1 = \Lambda_2 = \Lambda$
and $M_1 = M_2 = M$), but let $E \neq 0$. This means that the $Z_2$ symmetry is
broken only by the difference of the masses of black holes on the two sides of
the brane. After some redefinitions, in this case, from (\ref{cosmol}), we get
(again, for simplicity, taking $\epsilon_1 = \epsilon_2 = +1$)
\begin{equation}\label{dif bl hole masses}
m^4 X^4 - 4 M^6 X^3 - 4 M^6 \left( \frac{\rho + \sigma}{3m^2} -
\frac{\Lambda}{6} - \frac{C_1}{a^4} \right) X^2 + \frac{E^2}{m^4 a^8} = 0\, ,
\end{equation}
where $X$ stands for the expression $H^2 - (\rho+\sigma)/3m^2$. This result is
very similar to that obtained in Ref.~\refcite{KLM}, where this equation was
solved in the limit of small but nonzero $m$. In the limit $m \to 0$, equation
(\ref{dif bl hole masses}) reduces to the equation
\begin{equation}\label{dif bl hole masses no m}
H^2 = \frac{\Lambda}{6} + \frac{C_1}{a^4} + \frac{(\rho+\sigma)^2}{36 M^6} +
\frac{9E^2}{4 a^8(\rho + \sigma)^2}\, ,
\end{equation}
which was under consideration in Refs.~\refcite{CH,Ida,DVDP}.

\subsection{$m = 0$ and $\epsilon_1 = \epsilon_2 = +1$}

The limit $m=0$ was thoroughly investigated in the previous literature. Taking
this limit in Eq.~(\ref{cosmol}) and keeping $\epsilon_1=\epsilon_2=+1$, one
obtains
\begin{eqnarray}
H^2 &=& \frac{M_1^6\Lambda_1-M_2^6\Lambda_2}{6(M_1^6-M_2^6)} + \frac{E_2}{a^4}
+ \frac{M_1^6+M_2^6}{9(M_1^6-M_2^6)^2}
\left(\rho+\sigma\right)^2 \qquad\qquad\qquad\qquad\qquad\qquad \nonumber \\
&\pm& \frac{2M_1^3M_2^3}{9(M_1^6-M_2^6)^2} \left(\rho + \sigma\right) \left[
\left(\rho + \sigma\right)^2 + \frac{3}{2} (\Lambda_1-\Lambda_2) (M_1^6-M_2^6)
- \frac{C_2}{a^4} \right]^{1/2} \, , \label{no m}
\end{eqnarray}
where
\begin{equation}
C_2 = \frac{9}{4} \left[ \left( \frac{1}{M_2^6} - \frac{1}{M_1^6} \right) C - 3
E \left( \frac{1}{M_2^6} + \frac{1}{M_1^6} \right) \right] \, , \quad E_2 =
\frac{3E}{2(M_1^6-M_2^6)} \, .
\end{equation}

This equation was studied by Padilla.\cite{Padilla} He noticed that, in a
certain range of parameters, this model cosmologically behaves as the
$Z_2$-symmetric braneworld model with induced gravity. Specifically, with the
upper sign in (\ref{no m}) and with the assumptions
\begin{equation}\label{cond 1}
    \frac{\left| C_2 \right|}{a^4}\ll (\Lambda_1-\Lambda_2)(M_1^6-M_2^6)\, ,
\end{equation}
in which the right-hand side is assumed to be positive, and
\begin{equation}\label{cond 2}
    (\rho+\sigma)\ll
    \frac{M_1^3M_2^3}{M_1^6+M_2^6}\sqrt{(\Lambda_1-\Lambda_2)(M_1^6-M_2^6)} \, ,
\end{equation}
equation (\ref{no m}) reduces to
\begin{equation}\label{reduced no m}
H^2 = \frac{\rho+\sigma}{3m_{\rm eff}^2} + \frac{\Lambda_{\rm eff}}{6} +
\frac{E_2}{a^4} \, ,
\end{equation}
where
\begin{equation}\label{lambda eff}
\Lambda_{\rm eff} = \frac{M_1^6\Lambda_1-M_2^6\Lambda_2}{M_1^6-M_2^6} \, ,
\end{equation}
\begin{equation}\label{m eff}
m_{\rm eff}^2 = \sqrt{\frac{3\left(M_1^6-M_2^6\right)^3}
{2M_1^6M_2^6\left(\Lambda_1-\Lambda_2\right)}} \, ,
\end{equation}
which has the form of the Friedmann equation with cosmological constant and
``dark radiation'' [the last term in (\ref{reduced no m})].

\subsection{$M_1 = M_2$ and $\epsilon_1 = -\epsilon_2 = +1$}

The structure of Eq.~(\ref{cosmol}) allows one to consider the case $M_1 =
M_2=M$ with the (somewhat exotic) condition $\epsilon_1 = -\epsilon_2 = +1$. In
this case, we have
\begin{eqnarray}
m^4 \left( H^2  - \frac{\rho + \sigma}{3 m^2} \right)^2 &+& \frac{4M^{12}}{m^4}
\left[ \frac{H^2 - \left( \Lambda_1 + \Lambda_2 \right) /12
- E_1/12 a^4}{H^2 - (\rho + \sigma) / 3m^2 } \right]^2 \nonumber \\
&=& 4 M^6 \left( \frac{\Lambda_2 - \Lambda_1}{12} - \frac{C_1}{a^4} \right) \,
, \label{equal M different epsilon}
\end{eqnarray}
where $E_1$ is defined in (\ref{e1}), and $C_1$ in (\ref{equal lambda and M}).
In the limit $m \to 0$, this equation reduces to
\begin{equation}\label{equal M different epsilon no m}
H^2 = \frac{\Lambda_1 + \Lambda_2}{12} + \frac{E_1}{12 a^4} \pm
\frac{(\rho+\sigma)}{3M^3} \left[ \frac{\Lambda_2-\Lambda_1}{12} -
\frac{C_1}{a^4} - \frac{(\rho + \sigma)^2}{36 M^6} \right]^{1/2} \, .
\end{equation}
With the upper sign and for $\Lambda_2 \gg \Lambda_1$, this again has the form
of (\ref{reduced no m}) with different effective constants $m_{\rm eff}$ and
$\Lambda_{\rm eff}$.

\section{Conclusion}

In this letter, we derived a system of cosmological equations for a braneworld
which is a multivolume junction, i.e., a boundary of $N$ bulk spaces with
natural junction conditions, without assuming any symmetry between the bulk
spaces. In particular, the values of the fundamental constants, and even the
signatures of the extra dimension, could be different on different sides of the
brane. After that, we studied in more detail the usual case of two bulk spaces,
and obtained the closed system of scalar gravitational equations
(\ref{constraint-new})--(\ref{conserve-new}) describing the braneworld in the
absence of the $Z_2$ symmetry of reflection of the bulk with respect to the
brane, with different fundamental constants in the bulk, and even with possibly
different time signature of the fifth dimension on the two sides of the brane.
We then derived equation (\ref{cosmol}) describing the cosmological evolution
of such a brane and considered several special cases of this general situation.

\section*{Acknowledgments}

Yu.~S.\@ and A.~V.\@ were supported in part by the ``Cosmomicrophysics''
programme and by the Program of Fundamental Research of the Physics and
Astronomy Division of the National Academy of Sciences of Ukraine, by grant
No.~F16-457-2007 of the State Foundation of Fundamental Research of Ukraine,
and by the INTAS grant No.~05-1000008-7865. L.~N.~G.\@ acknowledges financial
support from COLCIENCIAS (Colombia).

\end{document}